\documentclass{npcs}
\usepackage{amsmath, amssymb}
\usepackage{graphics}
\usepackage{axodraw}
\usepackage{epsfig}
\usepackage{multicol}
\usepackage{amsmath}
\begin{document}

\runauthor{T.V. Shishkina, I.B. Marfin} \runtitle{Polarized lepton-nucleon scattering in
frame of various guage model}
\begin{topmatter}
\title{ Polarized lepton-nucleon scattering in
frame of various guage model}
\author{Author T.V. Shishkina}
\institution{NCHEP}
\address{153 Bogdanovitcha str.,220040 Minsk, Belarus}
\email{shishkina@hep.by}
\author{Author I.B. Marfin}
\institution{NCHEP}
\address{153 Bogdanovitcha str.,220040 Minsk, Belarus}
\email{marfin@hep.by}


\end{topmatter}
\newpage
\section{Inroduction}

$ $

The Standard theory of electroweak interactions is verified for a
long time. As a rule assumed that the Standard Model(SM) is only a
low energy effective theory, which will be able to resolve several
problems inherent to the Standard Model. Let us mention some of
these difficulties: 1) the scalar structure of the SM: nature of
the Higgs boson and origin of the electroweak symmetry breaking;
2) the huge number of SM free parameters to be fixed by
experiments; 3) the origin of parity violation for weak
interactions; 4) the origin of the SM three generation, etc.
Besides, the ultimate unification of all particles and of all
interactions is still an essential aim of particle physics. It is
natural to have several models or theories which go beyond the
Standard Model in order to satisfy this unification goal and to
resolve some problems of the SM like those mentioned above. A
common prediction to all these models is the existence of new Z'
bosons.

 The application of expanded gauge theories extends potentialities of
the electroweak theory. The predictions of this gauge models can
be conformed to experimental data with the widly changing
parametrs that are deviated from the Standard model. The
experiments of scattering of polarized leptons by unnpolarized
nucleon target when the transfer momentum runs up to $\sim
10^{40}$GeV${}^2$ and accordingly electroweak asymmetries are
closed to several decades of percent \cite{c1} allow to verify
predictions of different gauge models with an additional Z-boson.

There is important to consider the suplementary information about a
structure of one electoweak group using investigation of deep
inelastic scattering of polarized  leptons and
nucleons (see, for example, lepton-proton scattering at  HERA \cite{c2,c3,c28}).
The covariant method to calculate cross sections of processes
where two (and more) polarized particles are interacted has great 
importance. This method allows us to do  research 
not depending from concrete kinematics of different experiments. In
the paper  we have calculated the Lorentz-invariant differential
cross section of electroweak deep inelastic scattering of
polarized leptons by polarized nucleon target within the framework
of  gauge models with  additional Z-bozon.

\section{The  gauge models}
\subsection{The model \it{$SU(3)\times U(1)$}}

$ $

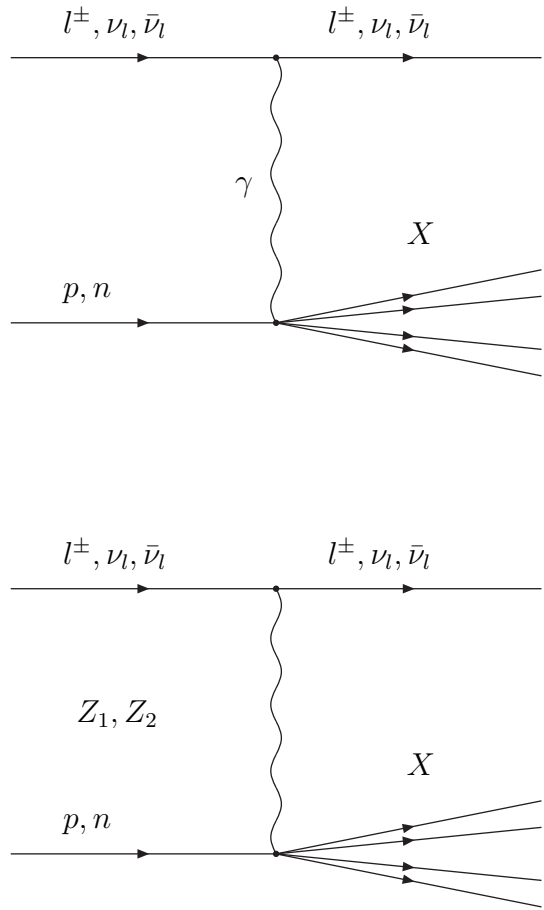
\begin{figure}[h]
\begin{picture}(600,200)(-70,0)
\ArrowLine(50,150)(150,150) \Photon(150,150)(150,50){2}{4}
\Vertex(150,150){1.2}
\ArrowLine(150,150)(250,150)\Vertex(150,50){1.2}
\ArrowLine(150,50)(250,30) \ArrowLine(50,50)(150,50)
\ArrowLine(150,50)(250,40) \ArrowLine(150,50)(250,70)
\ArrowLine(150,50)(250,60) \put(70,160){$l^{\pm}, \nu_{l},
\bar{\nu}_{l}$} \put(170,160){$l^{\pm}, \nu_{l}, \bar{\nu}_{l}$}
\put(135,100){$\gamma$}
\put(70,60){$p, n$} \put(200,80){$X$}
\end{picture}

\begin{picture}(600,200)(-70,0)\label{p2}
\ArrowLine(50,150)(150,150) \Photon(150,150)(150,50){2}{4}
\Vertex(150,150){1.2}
\ArrowLine(150,150)(250,150)\Vertex(150,50){1.2}
\ArrowLine(150,50)(250,30) \ArrowLine(50,50)(150,50)
\ArrowLine(150,50)(250,40) \ArrowLine(150,50)(250,70)
\ArrowLine(150,50)(250,60) \put(70,160){$l^{\pm}, \nu_{l},
\bar{\nu}_{l}$} \put(170,160){$l^{\pm}, \nu_{l}, \bar{\nu}_{l}$}
\put(75,100){$ Z_1, Z_2$}
\put(70,60){$p, n$} \put(200,80){$X$}
\end{picture}
\caption{The lepton-nucleon deep inelastic scattering}\label{p1}
\end{figure}

The interaction between a polarized lepton and polarized nucleon in
different gauge models can be described by following Feynmans` diagrams in fig. \ref{p1}

The electroweak interaction \cite{c5,c7} is
\begin{align}\label{a3}
L^{NC} & =
\frac{g\cos{\theta_W}}{\sqrt{3-4\sin{\theta_W}^2}}\{X_{\mu}
(\frac{1}{2}\bar{\nu_e}\gamma^{\mu}(1+\gamma_5)\nu_e \nonumber\\ &
- \frac{1}{2}\bar{e}\gamma^{\mu}(1+\gamma_5)e -
\bar{e}\gamma^{\mu}(1-\gamma_5)e  +
\frac{1}{2}\bar{u}\gamma^{\mu}(1+\gamma_5)u \nonumber\\ & -
\frac{1}{2}\bar{d}\gamma^{\mu}(1+\gamma_5)d -
\bar{d}\gamma^{\mu}(1-\gamma_5)d) \nonumber\\ & +
\frac{1}{2}\tan{\theta_W}^2
Z^{\mu}(\bar{\nu_e}\gamma^{\mu}(1+\gamma_5)\nu_e -
\bar{e}\gamma^{\mu}(1+\gamma_5)e\nonumber\\ & +
\bar{u}\gamma^{\mu}(1+\gamma_5)u -
\bar{d}\gamma^{\mu}(1+\gamma_5)d- 2J^{\mu}_{e.m.})\}.
\end{align}
$X^{\mu}, Z^{\mu}$ -- the neutral gauge bosons. The physical
bosons are obtained with following orthogonal transformations
\begin{align}\label{a4}
Z_1 = & \cos{\alpha}Z - \sin{\alpha}X, \nonumber\\ Z_2 = &
\sin{\alpha}Z + \cos{\alpha}X,
\end{align}
$$ -\pi/2 \leq \alpha \leq \pi/2.$$ Masses for $Z_1$ - and $ Z_2$ - bosons
are equal to
\begin{align}\label{a5}
M_1 =  80\mbox{GeV}, \nonumber\\ M_2 =  110 \mbox{GeV},
\nonumber\\ M_2/M_1  \gtrsim  1.4, \nonumber\\  \mbox{at $\alpha
= -0.02\pi$},
\end{align}
accordingly.
$$ \sin{\theta_W}^2 = 0.18 \sim 0.30. $$

\subsection{ The model \it{$SU(2)_L\times SU(2)_R\times U(1)$}}

$ $

In this model the electroweak interaction Langragian is defined
like the latter in $\it SU(3)\times U(1)$ (see (\ref{a3})).
\begin{align}\label{a6}
L^{NC} & =
\frac{g\cos{\theta_W}}{\sqrt{1-2\sin{\theta_W}^2}}\{X_{\mu}
(\frac{1}{2}\bar{\nu_e}\gamma^{\mu}(1-\gamma_5)\nu_e \nonumber\\ &
- \frac{1}{2}\bar{e}\gamma^{\mu}(1-\gamma_5)e  +
\frac{1}{2}\bar{u}\gamma^{\mu}(1-\gamma_5)u \nonumber\\ & -
\frac{1}{2}\bar{d}\gamma^{\mu}(1-\gamma_5)d)  \nonumber\\ & +
\frac{1}{2}\tan{\theta_W}^2
Z^{\mu}(\bar{\nu_e}\gamma^{\mu}(1+\gamma_5)\nu_e -
\bar{e}\gamma^{\mu}(1+\gamma_5)e\nonumber\\ & +
\bar{u}\gamma^{\mu}(1+\gamma_5)u -
\bar{d}\gamma^{\mu}(1+\gamma_5)d- 2J^{\mu}_{e.m.})\}.
\end{align}
Physical bosons masses are
\begin{align}\label{a7}
M_1 =  96\mbox{GeV}, \nonumber\\ M_2 =  288 \mbox{GeV},
\nonumber\\ M_2/M_1  \gtrsim  3.0, \nonumber\\  \mbox{at $\alpha
= -0.02\pi$}.
\end{align}

As one can see, the bosons become more massive in comparison with $\it SU(3)\times
U(1)$, besides that, the range of Weinberg`s angle is expanded --
$\sin{\theta_W}^2 = 0.16 \sim 0.34.$

\subsection{ The model \it{$SU(2)_L\times SU(2)_R\times U(1)_L \times
U(1)_R$}}

$ $

The model is more difficult for consideration because
it has three massive neutral gauge bosons \cite{c4,c25,c26}. We
will investigate the special case of transformating the $\it SU(2)_R\times U(1)_R$ to
the group $\it U(1)$ at low energy. In this model only two neutral massive boson exist. The
Lagrangian of electroweak interaction is expressed as
\begin{align}\label{a8}
L^{NC} & = g\sin{\theta_W}Z^{1'}_{\mu}(\bar{e}\gamma^{\mu}\gamma_5
e - \frac{2}{3}\bar{u}\gamma^{\mu}\gamma_5 u +
\frac{1}{3}\bar{d}\gamma^{\mu}\gamma_5 d)\nonumber\\ & +
\frac{g}{\sqrt{1-2\sin{\theta_W}^2}}Z^{2'}_{\mu}(\frac{1}{2}
\bar{\nu_e}\gamma^{\mu}(1+\gamma_5)\nu_e \nonumber\\ & -
\frac{1}{2} \bar{e}\gamma^{\mu}(1+\gamma_5)e + \frac{1}{2}
\bar{u}\gamma^{\mu}(1+\gamma_5)u\nonumber\\ & - \frac{1}{2}
\bar{d}\gamma^{\mu}(1+\gamma_5)d -
2\sin{\theta_W}^2\{-\bar{e}\gamma^{\mu}(1+\gamma_5)e\nonumber\\ &
+ \frac{2}{3}\bar{u}\gamma^{\mu}(1+\gamma_5)u -
\frac{1}{3}\bar{d}\gamma^{\mu}(1+\gamma_5)d \}).
\end{align}
The masses are defined by formulae
\begin{align}\label{a9}
\frac{M_W^2}{M_1^2}  = (1-2\sin{\theta_W}^2)(1 +
\sqrt{\frac{\sin{\theta_W}^2}{1-2\sin{\theta_W}^2}}\frac{\cos{\alpha}}{\sin{\alpha}}),\nonumber\\
\frac{M_W^2}{M_2^2}  = (1-2\sin{\theta_W}^2)(1 -
\sqrt{\frac{\sin{\theta_W}^2}{1-2\sin{\theta_W}^2}}\frac{\sin{\alpha}}{\cos{\alpha}}).
\end{align}
Using (\ref{a9}) one can obtain that boson masses are
\begin{align}\label{a10}
M_1 =  98\mbox{GeV}, \nonumber\\ M_2 =  181 \mbox{GeV},
\nonumber\\
 \mbox{at $\alpha
= -0.02\pi$}.
\end{align}

\section{The deep inelastic scattering of polarized leptons on polarized nucleons. Born approximation}

$ $

The expression for differential cross section \cite{c4} of longitudinally-polarized
leptons (with helicity $P_l$) in deep inelastic scattering (DIS) on polarized nucleons
are given by
\begin{align}\label{a11}
\sigma^{l^{\pm}}(k_1,k_2) & =
\frac{2\alpha^2Mx}{SQ^6}\sum_{q}f^{+}_q(x)\sum_{i=1,2}\{\frac{1}{2}f_q^2I_1-2f_qr_i
 (v_q^iG_V^{i\pm}I_1 + a_q^iG_A^{i\pm}I_2)+\nonumber\\ & +
\sum_{j=1,2}r_ir_j[(v_q^iv_q^j  +
a_q^ia_q^j)P_V^{ij\pm}I_1 + (v_q^ia_q^j +
a_q^iv_q^j)P_A^{ij\pm}I_2 ]\}\nonumber\\ &  -
\frac{4\alpha^2Mx}{SQ^6}\sum_{q}f^{-}_q(x)\sum_{i=1,2}\{\frac{1}{2}f_q^2P_lQ^2(K\eta)
 + 2r_if_q(a_q^iG_V^{i\pm}xI_1^{\eta} +
v_q^iG_A^{i\pm}(K\eta))\nonumber\\ & -
\sum_{j=1,2}r_ir_j[(v_q^ia_q^j +
a_q^iv_q^j)P_V^{ij\pm}xI_1^{\eta} +
(v_q^iv_q^ja_q^ia_q^j)P_A^{ij\pm}Q^2(K\eta) - \nonumber\\ & -
2a_q^ia_q^jP_A^{ij\pm}xI_2^{\eta} +
2x\varepsilon(\hat{p_1}\hat{\eta}\hat{k_1}\hat{k_2})(a_q^iv_q^j
-v_q^ia_q^j)]\},
\end{align}
where we use the definitions $$\sigma(k_1,k_2) = k_2^0 d^3\sigma/d^3 K_2, \,\,\, \alpha
=e^2/4\pi,$$ $$r_i =
Q^2G_i/[4(Q^2+M^2_i)\sin{\theta_W}^2(1-\sin{\theta_W}^2)],$$
$$G_V^{ij\pm} = g_V^i\pm P_lg_A^i,\,\,\, G_A^{ij\pm} = \mp g_A^i-
P_lg_V^i, $$ $$ P_V^{ij\pm} = g_V^ig_V^j +  g_A^ig_A^j \pm P_l(
g_V^ig_A^j +  g_A^ig_V^j),$$ $$ P_A^{ij\pm} = \mp(g_V^ig_A^j +
g_A^ig_V^j) - P_l( g_V^ig_V^j +  g_A^ig_A^j),$$ $$I_{1,2} = S^2
\pm X^2,\,\,\, I_{1,2}^{\eta} = S(k_{1,2}\eta)\pm X(k_{2,1}\eta),
$$ $$ K = k_1 + k_2,\,\,\, Q = k_1 - k_2,\,\,\, S = -2(p_1k_1),$$
$$ X = -2(p_1k_2),\,\,\, x=Q^2/(S-X),\,\,\, f_q^{\pm} = h^{+}_q(x)
\pm h^{-}_q(x), $$ $$
\varepsilon(\hat{p_1}\hat{\eta}\hat{k_1}\hat{k_2}) =
\varepsilon_{\alpha\beta\rho\sigma}p_{1\alpha}\eta_{\beta}k_{1\rho}k_{2\sigma};$$
$(h_q^{-})h_q^{+}$ is distribution functions of q-th quark that has a (anti)parallel to polarization of nucleon
polarization, $f_q$ is
a charge q-th quark, $k_1,\,p_1,\,(k_2,\,p_2)$ is 4-momenta for
an initial (final) lepton and nucleon, $\eta$ is 4-momentum of
initial nucleon polarization, M -- the nucleon mass, $M_i$ -- the
$Z_i$-boson mass, $g_A^i,\,g_V^i,\,v_q^i,\,a_q^i$ -- parameters of
discussed models. The last are  given in tables 1 and 2.

The covariant expression for the differential cross sections of
deep inelastic scattering within the framework of the nonminimal
gauge models of electroweak interaction  depends not only on
kinematics variables, but on polarizations of interaction
particles also. By means of changing of these variables one can
obtain electroweak asymmetries which are determined by the universal
formula:
\begin{eqnarray}\label{b1}
A^{e\omega}(P_{l_{1}},P_{N_{1}},e_{f_{1}},P_{l_{2}},P_{N_{2}},e_{f_{2}})=
\frac{\sigma^{e_{f_{1}}}(P_{l_{1}},P_{N_{1}})-\sigma^{e_{f_{2}}}(P_{l_{2}},P_{N_{2}})}{\sigma^{e_{f_{1}}}(P_{l_{1}},P_{N_{1}})+\sigma^{e_{f_{2}}}(P_{l_{2}},P_{N_{2}})}.
\end{eqnarray}
Here $\sigma^{e_{f}}(P_{l},P_{N})$ is the differential cross
section of scattering of  leptons with helicity
$P_{l}$ by a nucleon with longitudinal polarization  $P_{N}$.
4-momentum of the nucleon`s polarization is
$\eta=P_{N}(\vec{k_1}/E,0)$, where $\vec{k_1}$ -- is momentum of a
scattering nucleon, $E$ -- its energy, $e_f$ - lepton`s electric
charge. Eq.(\ref{b1}) defines all asymmetries based on a bunch
of the polarized particles, on a polarized target, and also on a
study of interaction of two polarized particles. The last item
 allows us to build up the most extended
succesion of electroweak asymmetries:

\noindent polarized asymmetries
\begin{eqnarray}\label{b2}
{A_p}^{\pm}=\frac{\sigma^{e\pm}(1,1)-\sigma^{e\pm}(-1,-1)}{\sigma^{e\pm}(1,1)+\sigma^{e\pm}(-1,-1)},\,\,\nonumber\\
{A_a}^{\pm}=\frac{\sigma^{e\pm}(1,-1)-\sigma^{e\pm}(-1,1)}{\sigma^{e\pm}(1,-1)+\sigma^{e\pm}(-1,1)},\,\
\end{eqnarray}
charge-polarized asymmetries
\begin{eqnarray}\label{b3}
{B_p}^{\pm}=\frac{\sigma^{e\pm}(1,1)-\sigma^{e\mp}(-1,-1)}{\sigma^{e\pm}(1,1)+\sigma^{e\mp}(-1,-1)},\,\,\nonumber\\
{B_a}^{\pm}=\frac{\sigma^{e\pm}(1,-1)-\sigma^{e\mp}(-1,1)}{\sigma^{e\pm}(1,-1)+\sigma^{e\mp}(-1,1)},\,\
\end{eqnarray}
and charged asymmetries
\begin{eqnarray}\label{b4}
C_{RR}=\frac{\sigma^{e^{+}}(1,1)-\sigma^{e^{-}}(1,1)}{\sigma^{e^+}(1,1)+\sigma^{e^-}(1,1)},\,\,\
\ \ \ \ \ \ \ \  \nonumber\\
C_{LL}=\frac{\sigma^{e^+}(-1,-1)-\sigma^{e^-}(-1,-1)}{\sigma^{e^+}(-1,-1)+\sigma^{e^-}(-1,-1)},\,\,\nonumber\\
C_{RL}=\frac{\sigma^{e^+}(1,-1)-\sigma^{e^-}(1,-1)}{\sigma^{e^+}(1,-1)+\sigma^{e^-}(1,-1)},\,\,\
\ \ \ \  \\
C_{LR}=\frac{\sigma^{e^+}(-1,1)-\sigma^{e^-}(-1,1)}{\sigma^{e^+}(-1,1)+\sigma^{e^-}(-1,1)},\,\,\
\ \ \ \ \nonumber\
\end{eqnarray}
where $\sigma^{e\pm}(1,1)$, $\sigma^{e\pm}(1,-1)$,
$\sigma^{e\pm}(-1,1)$, $\sigma^{e\pm}(-1,-1)$ -- the  differential
cross section of electroweak deep inelastic scattering of
leptons by nucleons with following values of polarizations of
particles  in the process: $P_l=1,\,\,P_N=1$;
$P_l=1,\,\,P_N=-1$; $P_l=-1,\,\,P_N=1$; $P_l=-1,\,\,P_N=-1$.

The analysis of the electroweak asymmetries at $E = 10 - 40 $ GeV
in the ep-DIS and at $E = 100 - 2000 $ GeV in the $\mu$p-DIS in
the framework of the quark--parton model  was made to estimate the
magnitude and behavior of the electroweak asymmetries are
calculated using the different extended gauge models, and to
compare their values with experimental data.

\begin{figure}[h!]
 \leavevmode
\begin{minipage}[b]{.475\linewidth}
\centering
\includegraphics[width=\linewidth, height=3.8in, angle=0]{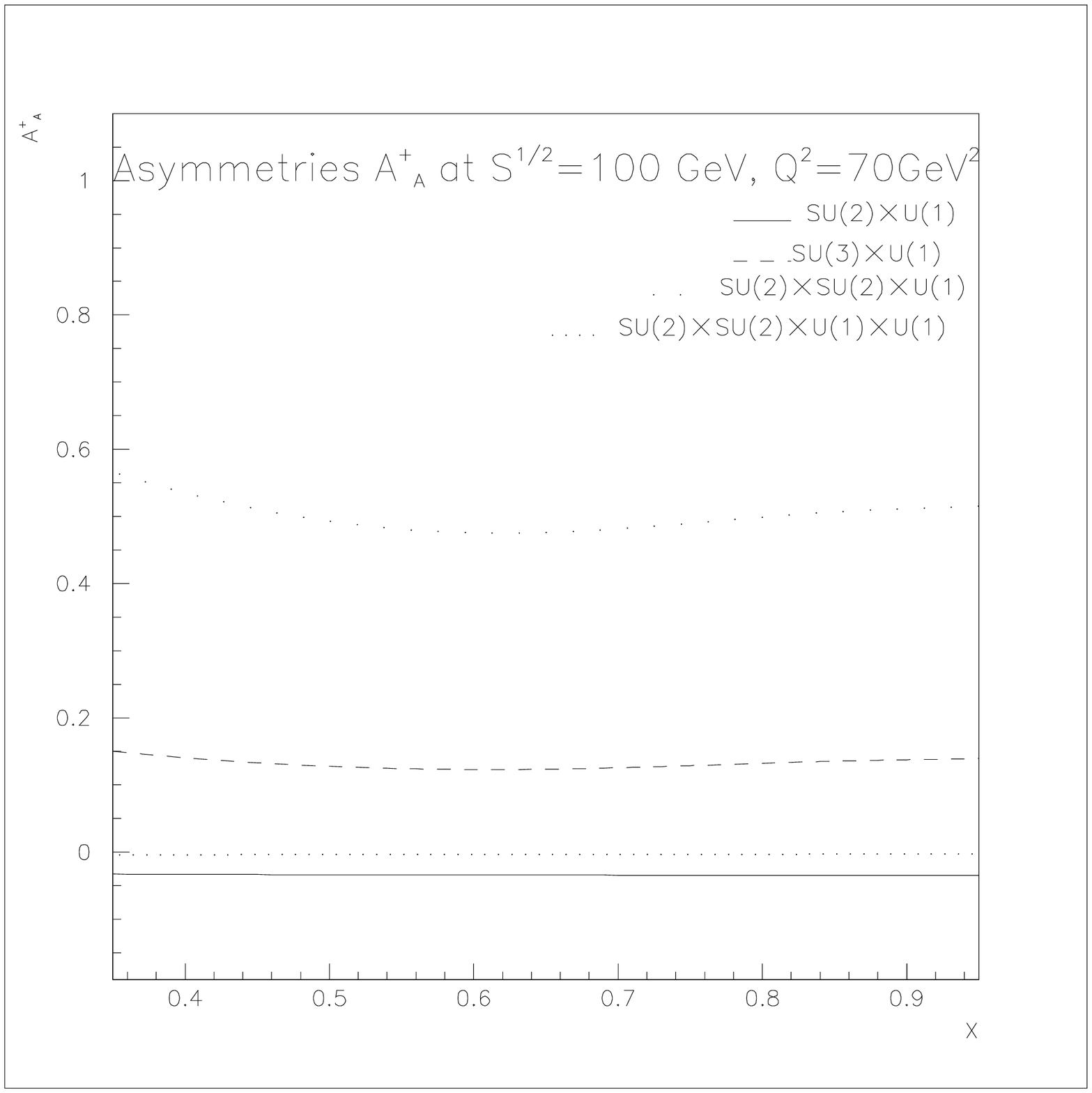}
\caption{The asymmetries $A^+_A$ in frame of various models.} \label{f1}
\end{minipage}\hfill
\begin{minipage}[b]{.475\linewidth}
\centering
\includegraphics[width=\linewidth, height=3.8in, angle=0]{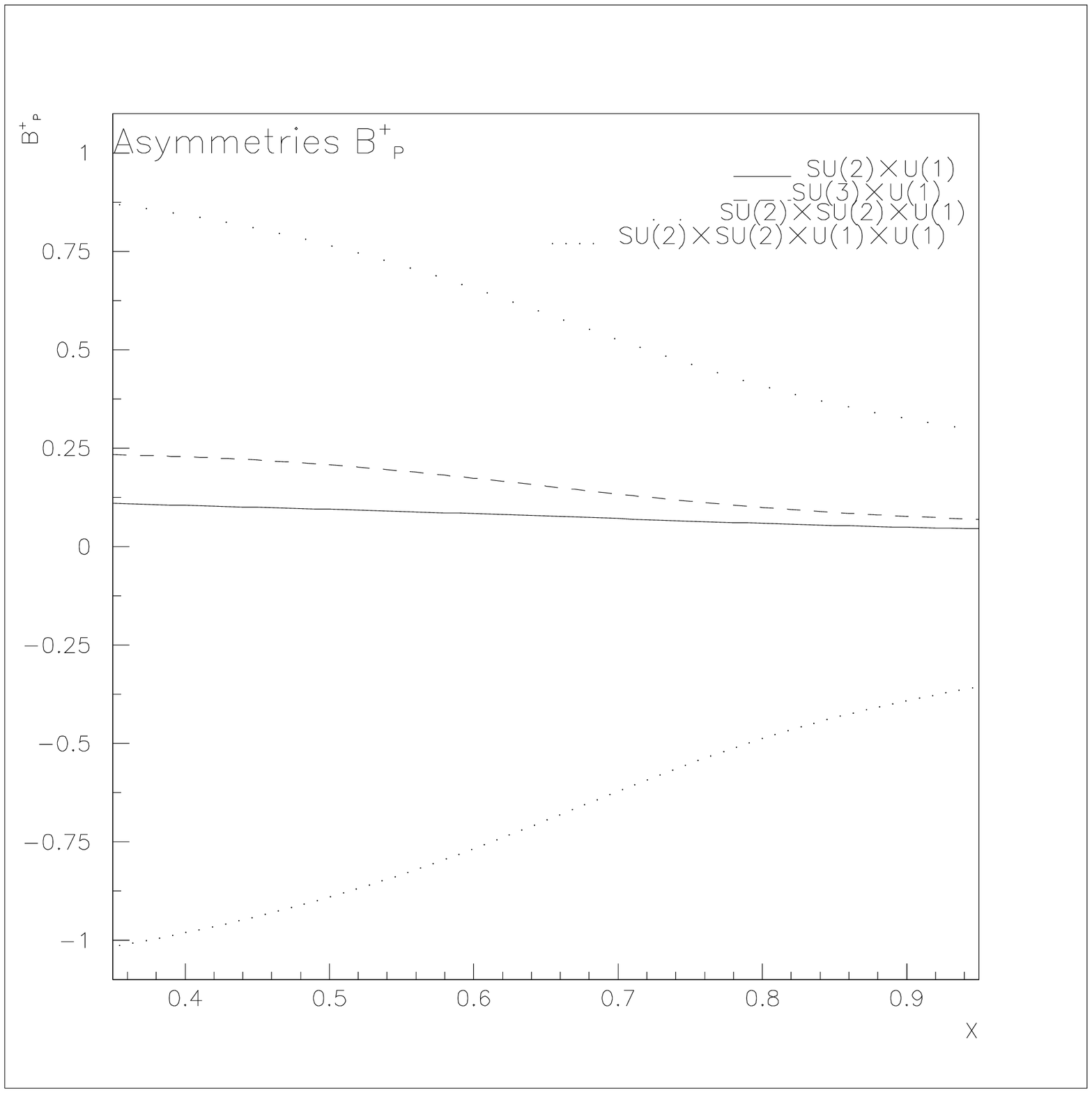}
\caption{The asymmetries $B^+_P$ in frame of various models.}
 \label{f2}
\end{minipage}
\end{figure}

\begin{figure}[h!]
 \leavevmode
\begin{minipage}[b]{.475\linewidth}
\centering
\includegraphics[width=\linewidth, height=3.8in, angle=0]{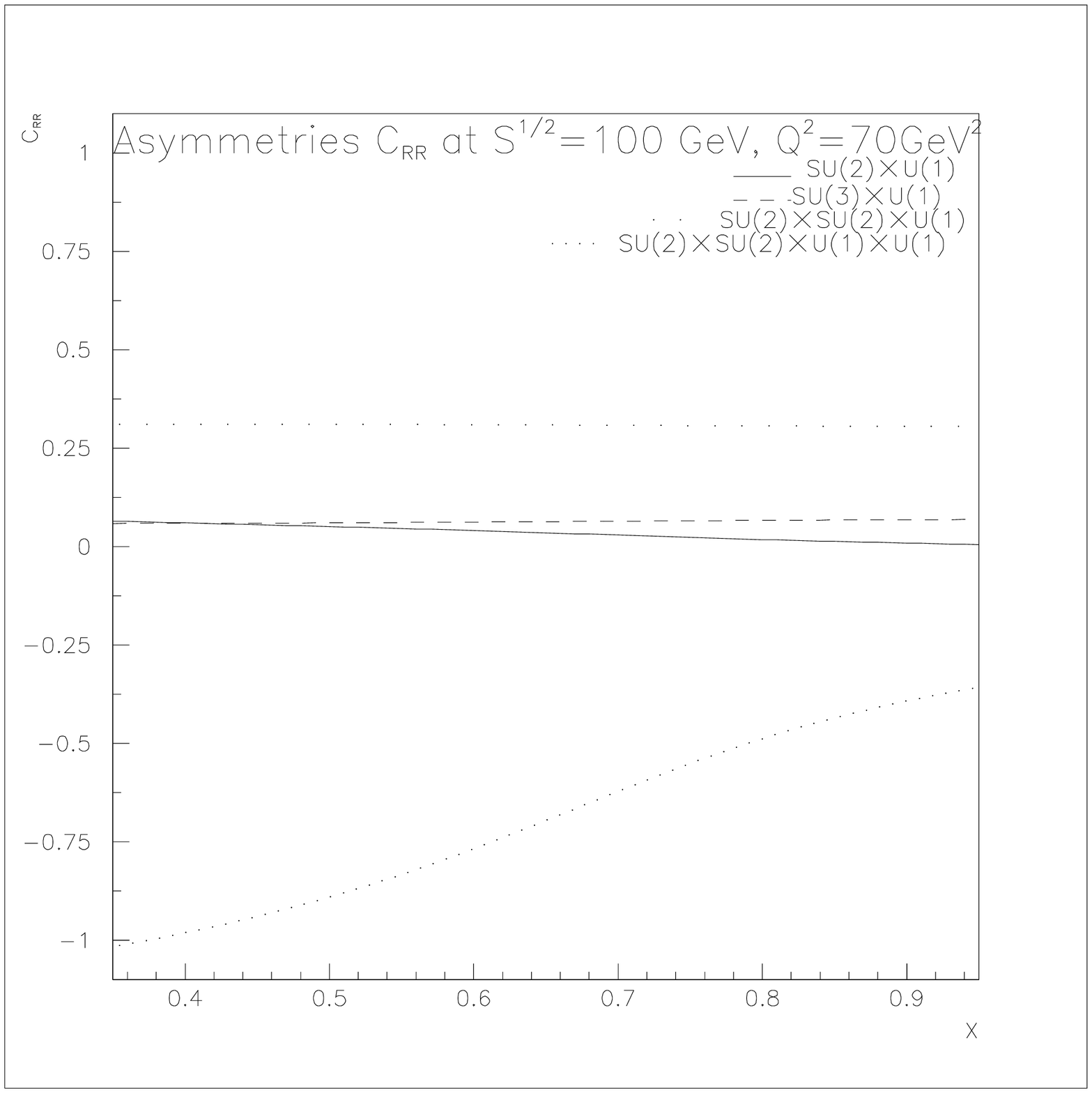}
\caption{The $C_{RR}$ asymmetries in frame of different models.}\label{f3}
\end{minipage}\hfill
\begin{minipage}[b]{.475\linewidth}
\centering
\includegraphics[width=\linewidth, height=3.8in, angle=0]{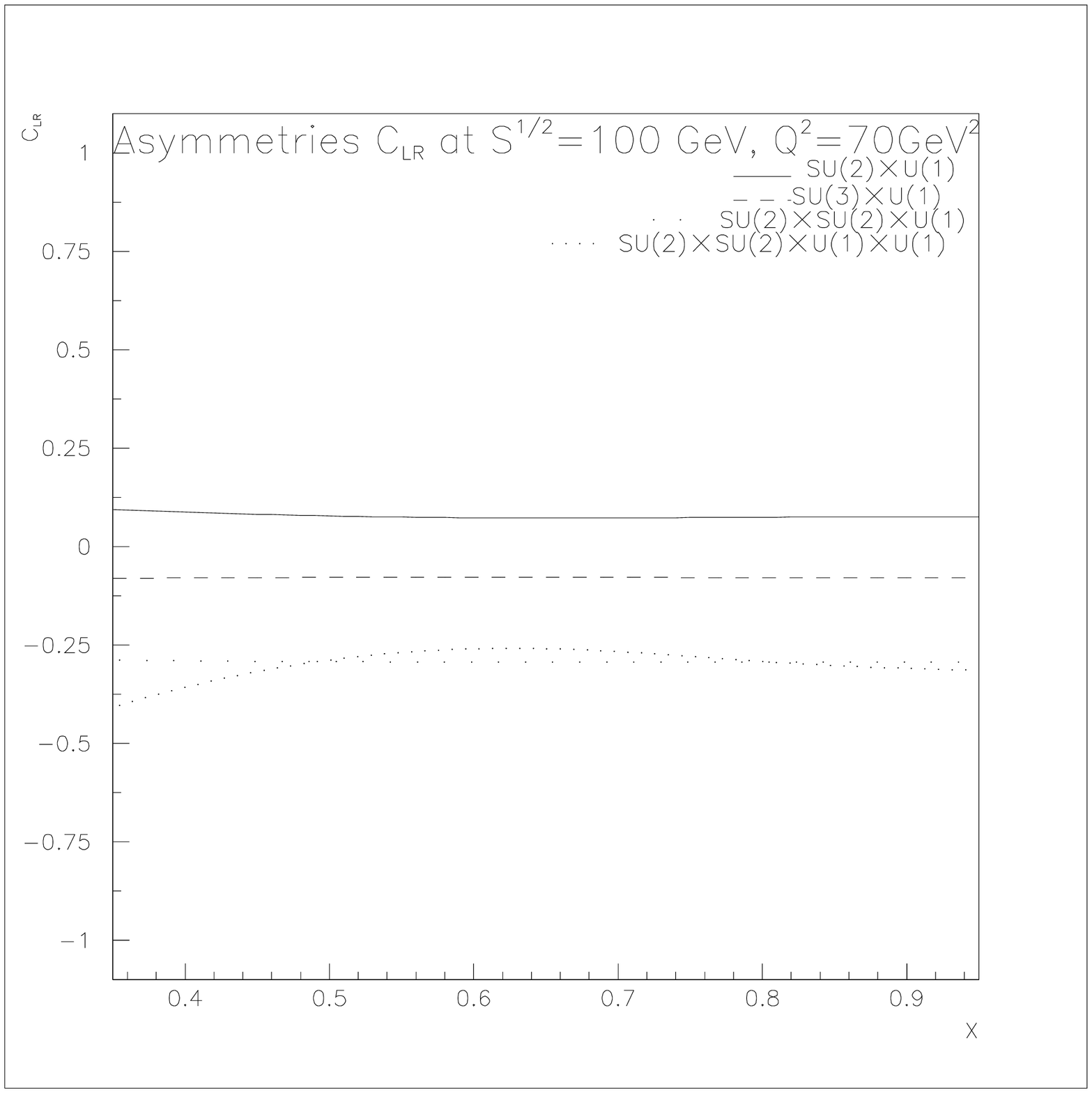}
\caption{The $C_{LR}$ asymmetries in frame of different models.}\label{f4}
\end{minipage}
\end{figure}

Some results of the electroweak asymmetries for the various
gauge models  are presented in figures \ref{f1} -- \ref{f4}. All asymmetries
are distinguished as functions of the  scailing variable $y$ at the fixed $x$. They are
equal to several percent at $E=100$ GeV and several tens of
percent as energy of scattered lepton goes up to $2000$ GeV and
they have trend to increase in the region of maximum $x$ and $y$. So
value of asymmetries $A^{\pm}_p$ reach $3\%$ at $E=100$ GeV and
$15\%$ at $E=1$ TeV but asymmetries $A^{\pm}_A$ -- $2\%$ and
$10\%$ accordingly. In the case of scattering of leptons with
positive and negative electric charge the asymmetries $A^{\pm}_p$
and $A^{\pm}_u$ have opposite signs, with $|A^{-}_p|$ exceeding
$|A^{+}_p|$ throughout kinematic region, and $|A^{+}_A|>|A^{-}_A|$.
Exceptions of these inequalities are results obtained in the
framework of the Standard Models of electroweak interaction where
$|A^{-}_A|>|A^{+}_A|$. The asymmetries significantly
depend on  longitudinal projections signs  of spins $P_l$
(in case of a lepton) and $P_N$ (in case of  a nucleon). Absolute values of
asymmetries with parallel combination of lepton and nucleon spins
exceed that of asymmetries $A^{\pm}_A$  over whole  kinematic region.
Comparison of different models` predictions for the electroweak
polarized asymmetries indicates the results obtained in the frame
of $\it SU(3)\times U(1)$ and $\it SU(2)_L \times SU(2)_R \times U(1)$
gauge models are closed to each other and greatly differ from
values of the asymmetries in other gauge models (see fig.\ref{f1},
fig.\ref{f2}).
It is particularly seen in the case of $\mu^{-}p$-DIS where they
can be two times differed in the region $y \geq 0.6$ and $x \leq
0.7$. The Standard Model has distinguished functional dependence
of asymmetries on  $y$. At the same time asymmetries
$A^{\pm}_A$ reach maximum values in the $\it SU(2)\times U(1)\times
[SU(2)]'$ model and $A^{\pm}_p$ -- in the $\it SU(3)\times U(1)$ one.
Replacement of scattering of a
positive muon by scattering of a negative muon leads to the great
change of the $y$-dependence of all asymmetries. It is
particularly evident in the $\it SU(2)\times U(1) \times [U(1)]'$
gauge model.

 The charge-polarized electroweak asymmetries $B^{\pm}_p$ and
$B^{\pm}_A$ are very sensitive to signs of longitudinal spins of
leptons and nucleons as well as asymmetries $A^{\pm}_p$ and
$A^{\pm}_A$ greatly depend on scaling variables. So $|B^{\pm}_A|$
  grows as  a linear function while $y$ increases, but
$|B^{\pm}_p|$ falls with $y$ decreasing, and then, changes its
sign and begins grow. It again approaches the minimum value at
$y\rightarrow 0$. The absolute value of $B^{\pm}_p$ exceeds that
of $B^{\pm}_u$ in the region of $x\leq 0.7$ and $y\geq 0.8$.

In the case of $\mu^{+}p$- and $\mu^{-}p$-DIS throughout
kinematic region  the asymmetries $B_p$ and $B_A$ nearly
identically depend on $y$, but they have opposite signs. Values of
asymmetries in the case of $\mu^{+}p$--DIS exceed values of
asymmetries in the case of $\mu^{-}p$--DIS.  In the case of $\mu^{+}p$--DIS
the asymmetries obtained in the discussed gauge models differ from
results obtained in the framework of the Standard Model greatly.
This difference particularly increases at high energy of
scattering particles and in the region where $x$ and $y$ are maximal.

Behavior of the charge electroweak asymmetries
 for scattering of particles with the
parallel spin configuration $C_{RR},C_{LL} $ are similar to behavior according to
the antiparallel spin configuration $C_{RL},C_{LR}$. The asymmetries for the
parallel spin configuration ($C_{RR}, C_{LL}$) decrease while $y$
fall to zero then they become negative and begin increase. These
asymmetries are minimal at $y\rightarrow 0$. The charge
asymmetries for the antiparallel spin configuration are almost
linear functions of $y$. In the case of $P_l\leq 0$ the charge
electroweak asymmetries $C_{LL}$, $C_{LR}$ greatly exceed values
of the asymmetries $C_{RR}$ and $C_{RL}$ in the whole kinematic
region (for instance, $C_{LL}$ -- 6 and 35 $\%$, $C_{LR}$ -- 4 and
20 $\%$ at lepton`s energy $E=100$ and $1000$ GeV accordingly),
but $C_{RL}$ and $C_{RR}$ increase from 2  to 12 $\%$ at these
values of energy.

The charge asymmetries depend on the gauge model choice (see
fig.\ref{f3}, fig.\ref{f4}).
The predictions of $\it SU(3)\times U(1)$ and $\it SU(2)_L \times
SU(2)_R\times U(1)$ are close, and the difference of their
asymmetries from asymmetries in the Standard Model is several tens
of percent. Exception is the  asymmetry $C_{RR}$, which has region
of values $0.3 \leq y \leq 0.6 $ for this difference, the
asymmetry $C_{RL}$ also has  close values in the framework of
nonminimal gauge models. The charge asymmetries reach the maximum
in the models $\it SU(2)\times U(1)\times [U(1)]'$ and $\it SU(2)\times
U(1)\times [SU(2)]'$ with result for all alternative models
(except $\it SU(3)\times U(1)$ and $\it SU(2)_L \times SU(2)_R\times
U(1)$) greatly differing from each other. This difference
surpasses $10\%$ in the significant part of the kinematic region.

All noted characteristics of the asymmetries behavior are kept in
case of scattered electrons (positrons) with energy up to $40$
GeV, although their values decrease about ten times.

Thus, the analysis of the asymmetries (\ref{b2}) -- (\ref{b4})
allows to discriminate gauge models of electroweak
interaction. The experimental research of gauge models are
complicated because of the fact that the differences between these
gauge models have significant values  near the border of the
kinematic region. Nevertheless the analysis of all possible
corrections to cross sections and asymmetries giving the basis for
investigation of gauge models, allowed to definite the complementary
asymmetries:
\begin{eqnarray}\label{b5}
{D_1}^\pm=\frac{\sigma^{e\pm}(0,1)-\sigma^{e\pm}(0,-1)}{\sigma^{e\pm}(1,1)-\sigma^{e\pm}(1,-1)},\,\,\
\ \ \ \ \nonumber\\
{D_2}^\pm=\frac{\sigma^{e\pm}(0,1)-\sigma^{e\pm}(0,-1)}{\sigma^{e\pm}(-1,1)-\sigma^{e\pm}(-1,-1)},\,\,\nonumber\\
{F_1}^\pm=\frac{\sigma^{e\pm}(1,0)-\sigma^{e\pm}(-1,0)}{\sigma^{e\pm}(1,1)-\sigma^{e\pm}(-1,1)},\,\,\
\ \ \ \ \\
{F_2}^\pm=\frac{\sigma^{e\pm}(1,0)-\sigma^{e\pm}(-1,0)}{\sigma^{e\pm}(1,-1)-\sigma^{e\pm}(-1,-1)},\,\
\nonumber
\end{eqnarray}
which give the greatly distinguishing values in the large part of
the kinematic region.
Their values  are differed significantly from that of
the other  models  mentioned above throughout kinematic region as well.
The results of the  models $\it SU(2)_L\times SU(2)_R\times U(1)$ and $\it SU(3)\times
U(1)$ are very close to each other.  All asymmetries significantly
decrease at high $\sqrt{S}$.

The above mentioned features of electroweak asymmetries give a
chance to check the Standard theory of electroweak interaction in
processes of deep inelastic scattering of polarized particles. The
asymmetries at high energy in different models are substaintially
distinguished, and that allows to observe difference between
nonminimal gauge models by experimental way.

\section{The radiative corrections to asymmetries of lN-deep inelastic scattering}

$ $

Radiative corrections can be sizeable in comparison
with a contribution of the additional gauge boson in the kinematic
region where electroweak asymmetries are only a few percent.  In the case of high energy radiative corrections to lN-DIS processes must be taken
into account. Therefore radiative effects are considered in the whole
kinematic region.

Electroweak asymmetries (\ref{b1}) with the lowest-order
 correction are presented in the following form:
\begin{eqnarray}\label{b6}
A^{e\omega}=\frac{{\sigma_0}^- + {\sigma_V}^- + {\sigma_R}^- +
{\sigma_{el}}^-}{{\sigma_0}^+ + {\sigma_V}^+ + {\sigma_R}^+ +
{\sigma_{el}}^+}.
\end{eqnarray}
Here $\sigma_0$, $\sigma_V$, $\sigma_R$, $\sigma_{el}$  are
contributions to the cross section of lN-scattering according
to a tree level cross section, exchange of an additional virtual
photon, radiative effects of the deep inelastic scattering and the
elastic peak. The signs +(-) indicate the sum (subtraction) of
cross sections of different spin configurations in asymmetries.
The electroweak correction to the asymmetry (\ref{b6}) of lp-DIS is 
expressed as 
$$\delta A =
A^{e\omega} - {A_0}^{e\omega} = \delta A_{in} + \delta A_{el},$$
where ${A_0}^{e\omega}$ is defined by equation (\ref{b1})). The main contribution in 
 wide kinematic region to the correction is given 
by $\delta A_{in}$. In frame of leading log approximation \cite{c25,c26} the radiative correction to asymmetry can
be obtained as
\begin{figure}[h!]
 \leavevmode
\begin{minipage}[b]{.475\linewidth}
\centering
\includegraphics[width=\linewidth, height=3.8in, angle=0]{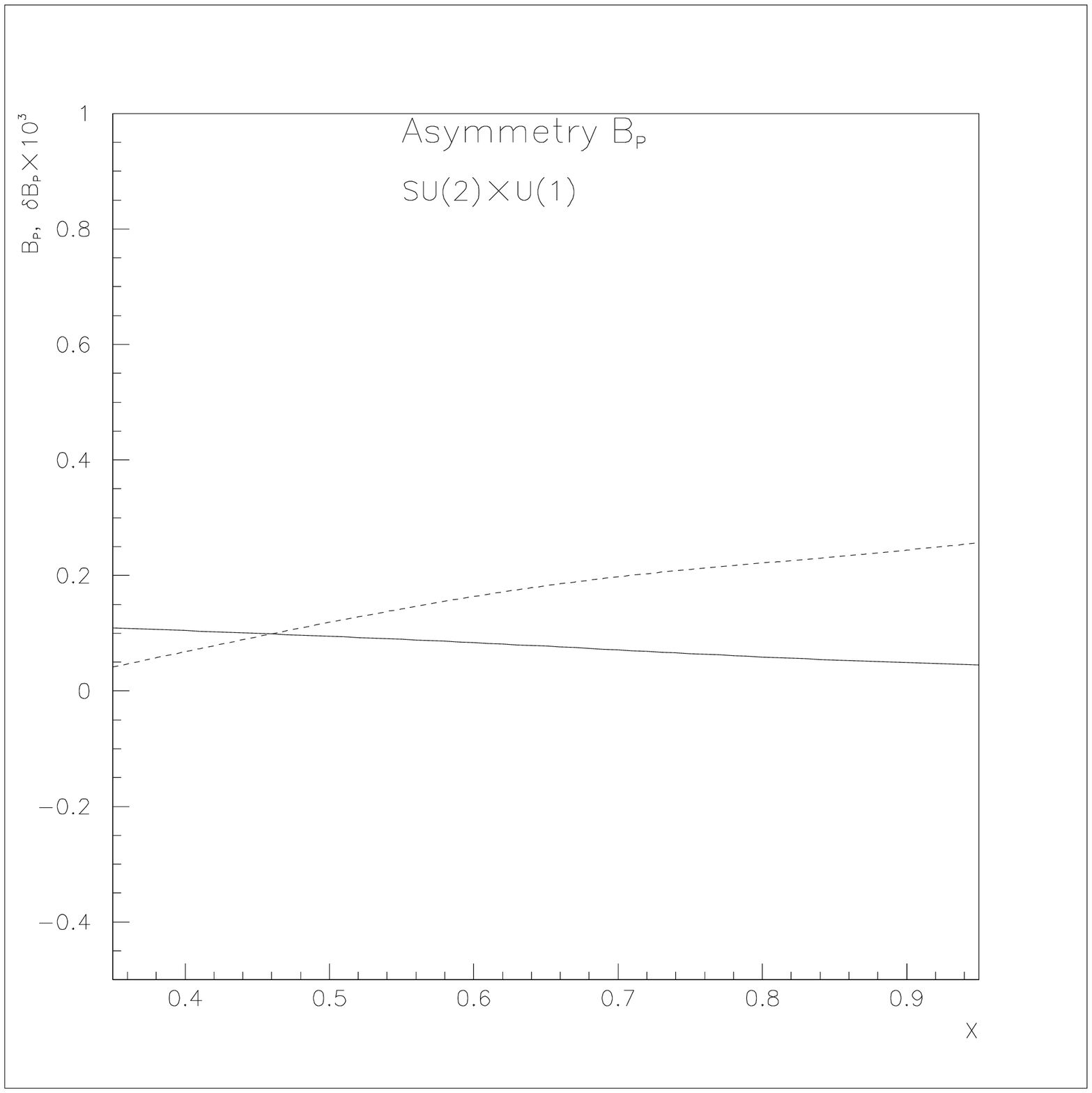}
\caption{The $B_P$ asymmetry in the Born approximation (solid line) and next to the Born approximation (dotted line)}\label{f5}
\end{minipage}\hfill
\begin{minipage}[b]{.475\linewidth}
\centering
\includegraphics[width=\linewidth, height=3.8in, angle=0]{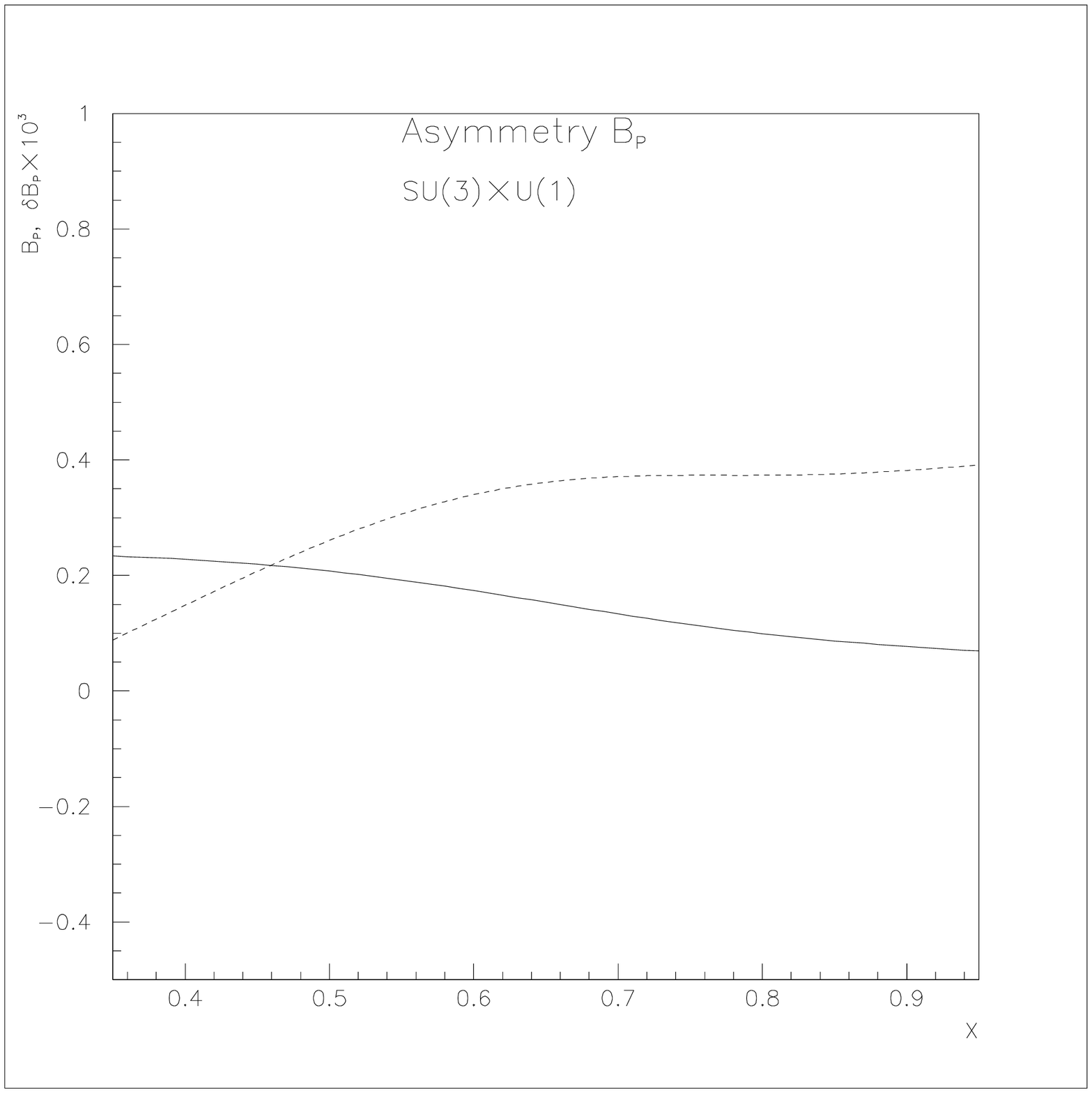}
\caption{The $B_{P}$ asymmetry in the Born approximation (solid line) and next to the Born approximation (dotted line) } \label{f6}
\end{minipage}
\end{figure}

\begin{eqnarray}\label{b7}
\begin{array}{c}
{\displaystyle \delta A_{in} =
\frac{\alpha}{2\pi}\ln(\frac{Q^2}{M^2})\sum_{i=1,2}{\int_{z_i}}^1
dz
\frac{1+z^2}{1-z}\frac{x_i{\sigma_0}^-(x_i,y_i,E_i)}{x\sigma_0(x,y,E)}\times}
\nonumber\\ {\displaystyle
\times[{A_0}^{e\omega}(x_i,y_i,E_i)-A_0(x,y,E)],}
\end{array}
\end{eqnarray}
where $$x_1 = z x_2, x_2 =( x y )/(z-1+y),$$ $$E_2 = E, E_1 =
zE,$$ $$z_1 = 1 - y - x y, z_2 = (1-y/(1 - x y),$$ $$y_1 =
y_2(z-1+y)/(z), Q^2 = (k_1 - k_2)^2.$$ Here the real photon radiation is also included.
 The consideration of diagrams with additional virtual photon leads to regularization of
$\sigma_{in}$ in the infrared region \cite{c27,c28}.
\begin{figure}[h!]
 \leavevmode
\begin{minipage}[b]{.475\linewidth}
\centering
\includegraphics[width=\linewidth, height=3.8in, angle=0]{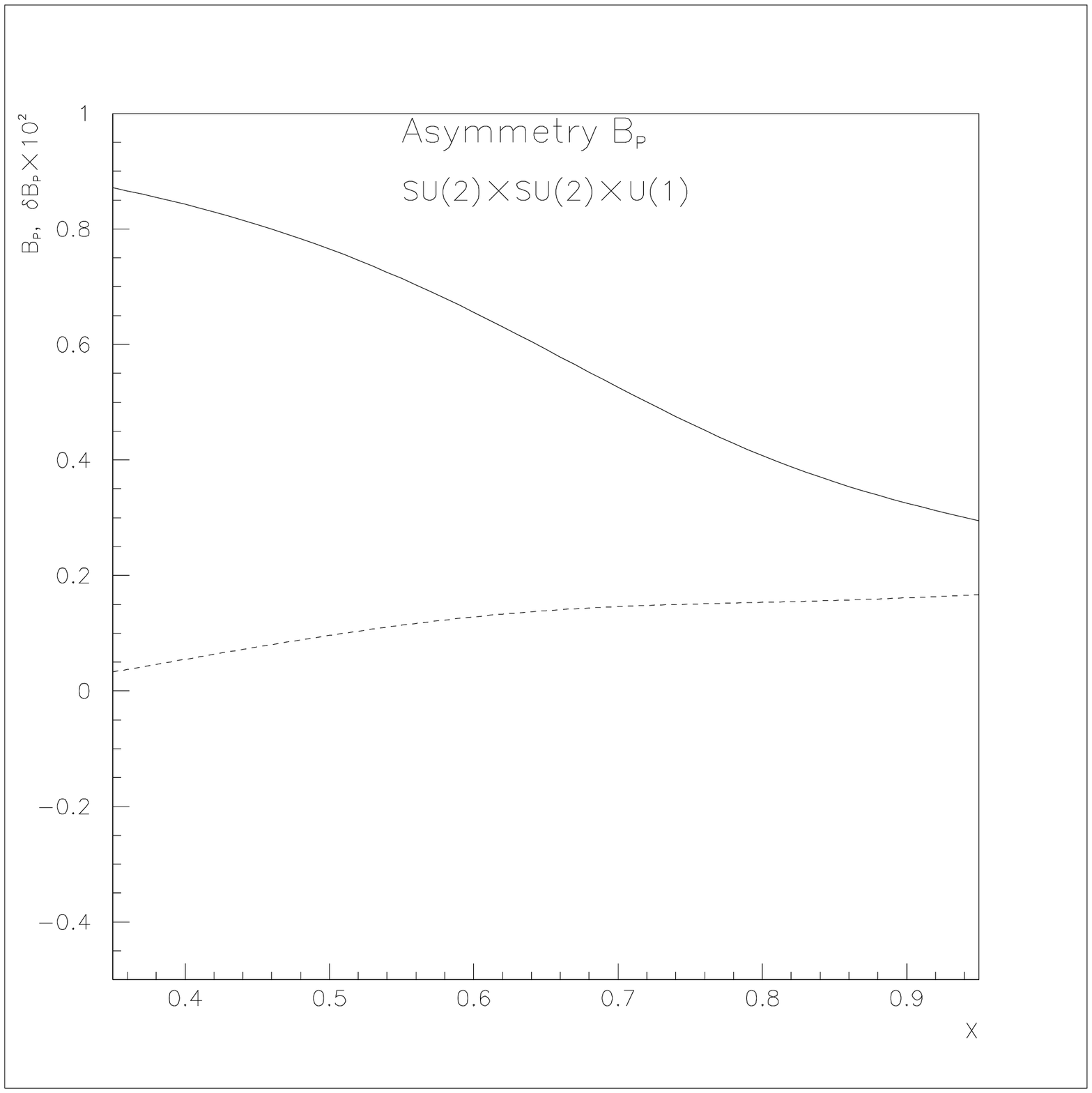}
\caption{The $B_{P}$ asymmetry in the Born approximation (solid line) and next to the Born approximation (dotted line)} \label{f7}
\end{minipage}\hfill
\begin{minipage}[b]{.475\linewidth}
\centering
\includegraphics[width=\linewidth, height=3.8in, angle=0]{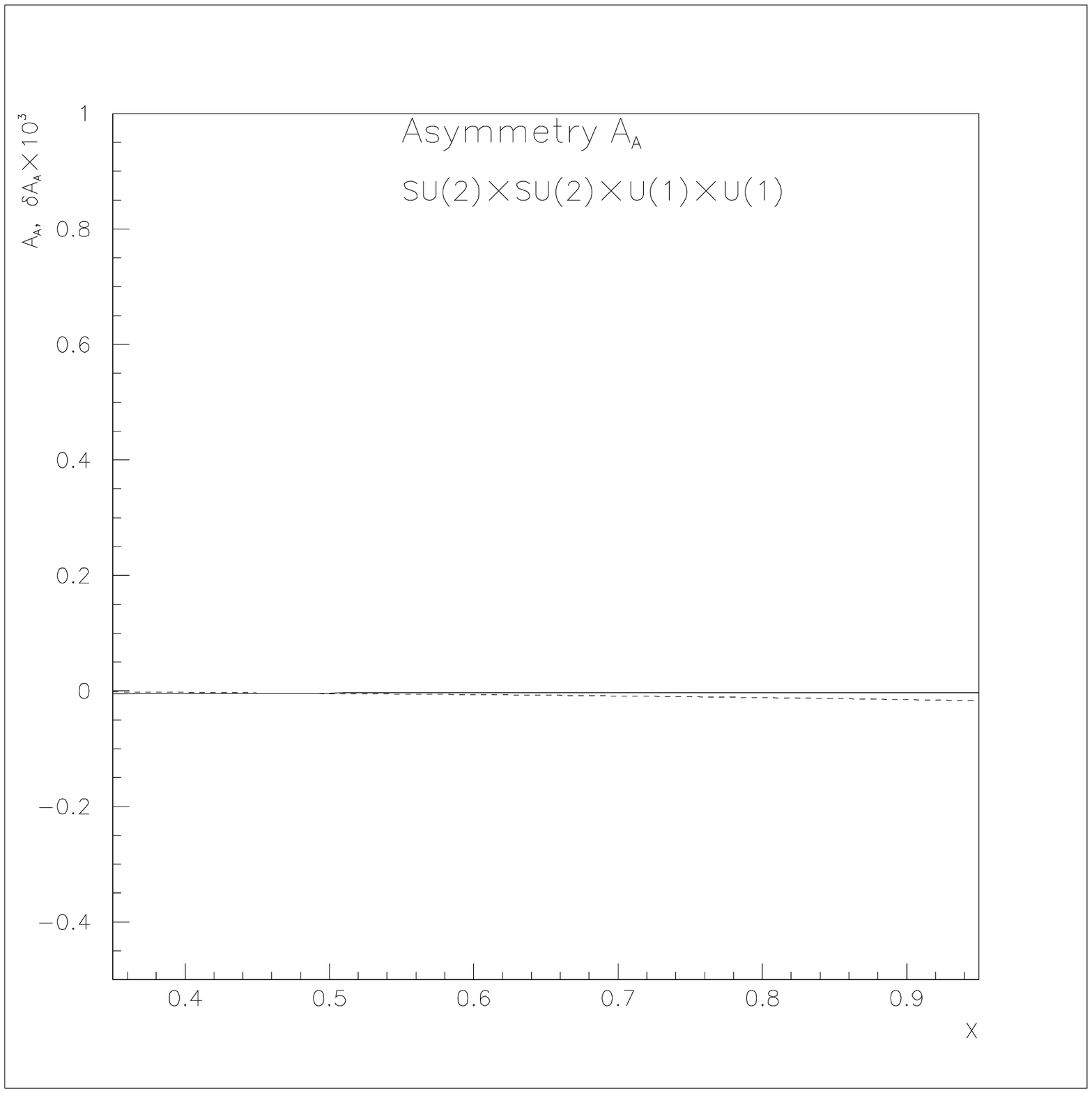}
\caption{The $A_{A}$ asymmetry in the Born approximation (solid line) and next to the Born approximation (dotted line)}\label{f8}
\end{minipage}
\end{figure}
 Within the limits of our approach the radiative elastic peak gives a insignificant
contribution stipulated by  the exchange of Z-bozon which is
proportional  to the bozon propagator under conditions:
$$m^2,\,\,M^2,\,\,t=(p_1-p_2)^2 << s=-2p_1k_1,\,\,X=
-2p_1k_2,\,\,{M_{Z_i}}^2,$$ where $m,\,\,M,\,\,M_{Z_1(Z_2)}$ --
lepton, nucleon, $Z_1(Z_2)$-bozons  masses, $p_1,\,\,p_2$ --
4-momenta of initial, final nucleons, $k_1,\,\,k_2$ - 4-momenta
of initial, final leptons. The electromagnetic correction
predominate over weak one. In this case  the contribution
${\sigma_{el}^-}$ is proportional to  $(P_{l_1} P_{N_1} - P_{l_2}
P_{N_2})$ that dismiss the electromagnetic correction in the
asymmetry. The cross section  ${\sigma_{el}^+}$ are defined by
cross section of elasic scattering averaged on the spins of
interaction particles. The most compact form of the differential
cross section in approach of leading log is:
\begin{eqnarray}\label{b8}
 \frac{d\sigma}{d x d y} =
 \frac{\alpha^3}{2}\frac{1+(1-y)^2}{1-y}f[\frac{x^2}{4(1-x)}],\,\
 \end{eqnarray}
 where $$f(x) = \int\limits_x^\infty
 d\tau[2(A_1(\tau))/(\tau)-(A_2(\tau))/(\tau^2)-(A_2(\tau))/(x\tau)],$$
 $A_1(\tau),\,\,A_2(\tau)$
 are expressed by electromagnetic nucleon form-factors $G_E$, $G_M$ as
 $$ A_1(t) =
{G_M}^2(t),\,\,A_2(t)=({G_E}^2(t)+\tau {G_M}^2(t))/(1+t),\,\,\tau
= t/4 M^2.$$

\begin{figure}[h!]
 \leavevmode
\begin{minipage}[b]{.475\linewidth}
\centering
\includegraphics[width=\linewidth, height=3.8in, angle=0]{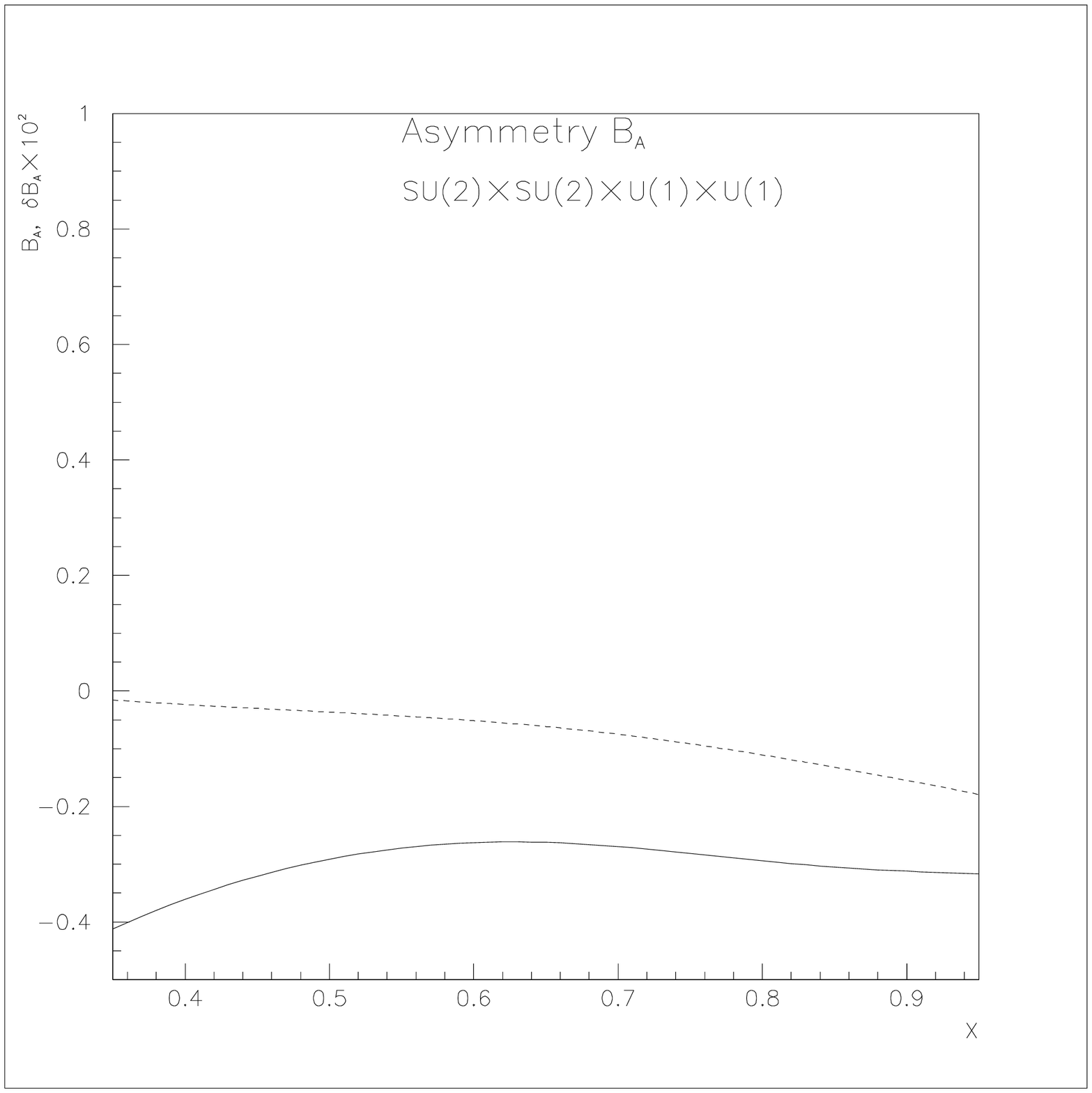}
\caption{The $B_{A}$ asymmetry in the Born approximation (solid line) and next to the Born approximation (dotted line)}\label{f11}
\end{minipage}\hfill
\begin{minipage}[b]{.475\linewidth}
\includegraphics[width=\linewidth, height=3.8in, angle=0]{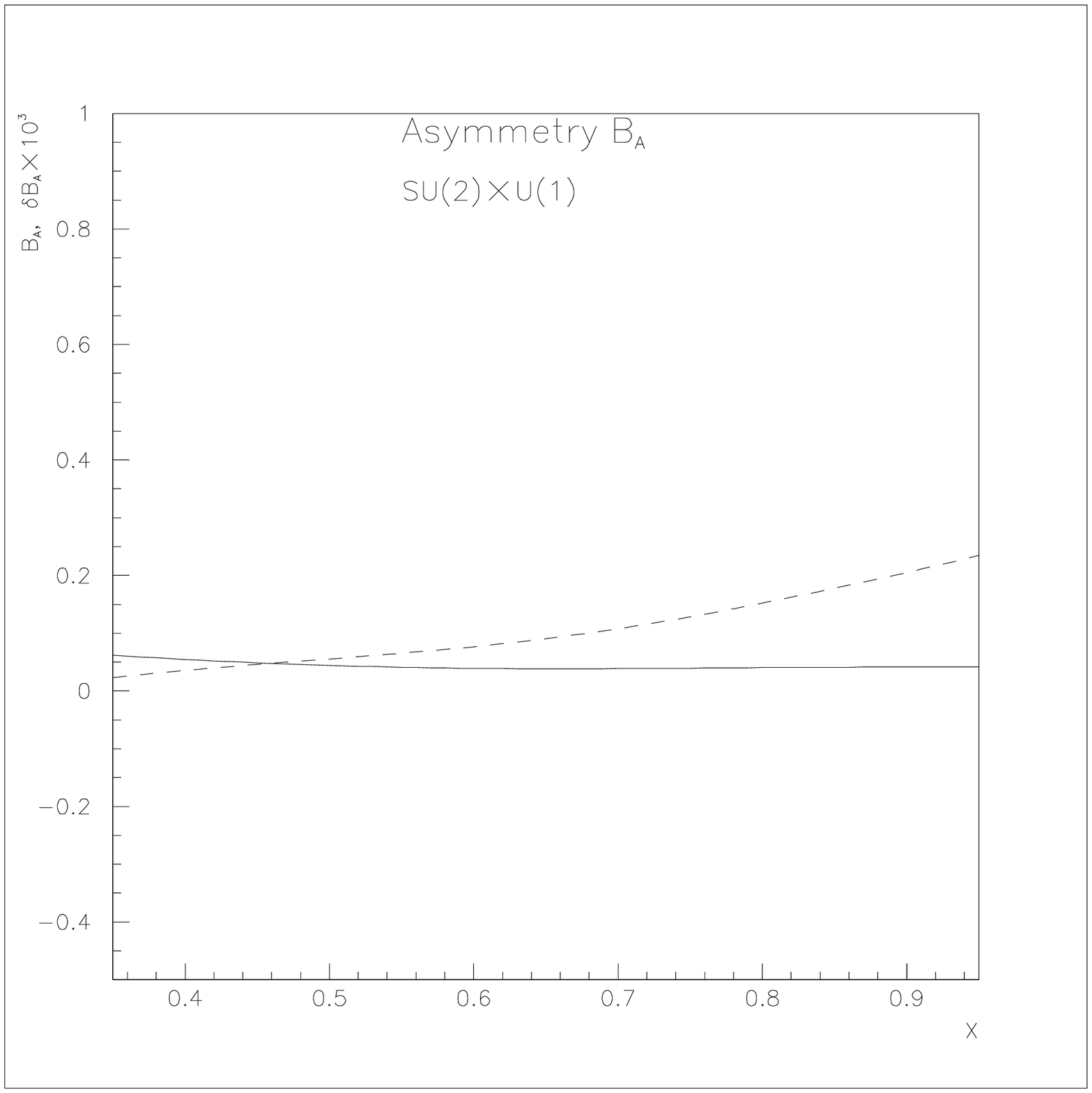}
\caption{The $B_A$ asymmetry in the Born approximation (solid line) and next to the Born approximation (dotted line)} \label{f12}
\end{minipage}
\end{figure}

The investigation of radiative effects (\ref{b6}) --(\ref{b7}) at
high energy near  $2000$ GeV discovers that  they are similar to
asymmetries of the polarized lepton-nucleon collison. The main
contribution into considered asymmetries of radiative correction
consists of two parts: electroweak and elastic scattering
corrections. The electroweak contribution is conditioned by the
radiation of real photons. It has maximum about ten percent in the
region  $x \leq 0.3,$ $y \geq 0.8$ and increases at high energy
and high transfer momentum. The radiative correction (\ref{b6}) by
$x\rightarrow 0$, where electroweak asymmetries are small, has a
large value. Also the elastic scattering correction reaches at a
value about tens percent on $y \geq 0.7$. The leading log
approach (\ref{b8}) gives about 90\% of the correction  $\delta
A_{el}$ to electroweak asymmetries (see fig.\ref{f5} -- fig.\ref{f12}).
\begin{figure}[h!]
 \leavevmode
\begin{minipage}[b]{.475\linewidth}
\centering
\includegraphics[width=\linewidth, height=3.8in, angle=0]{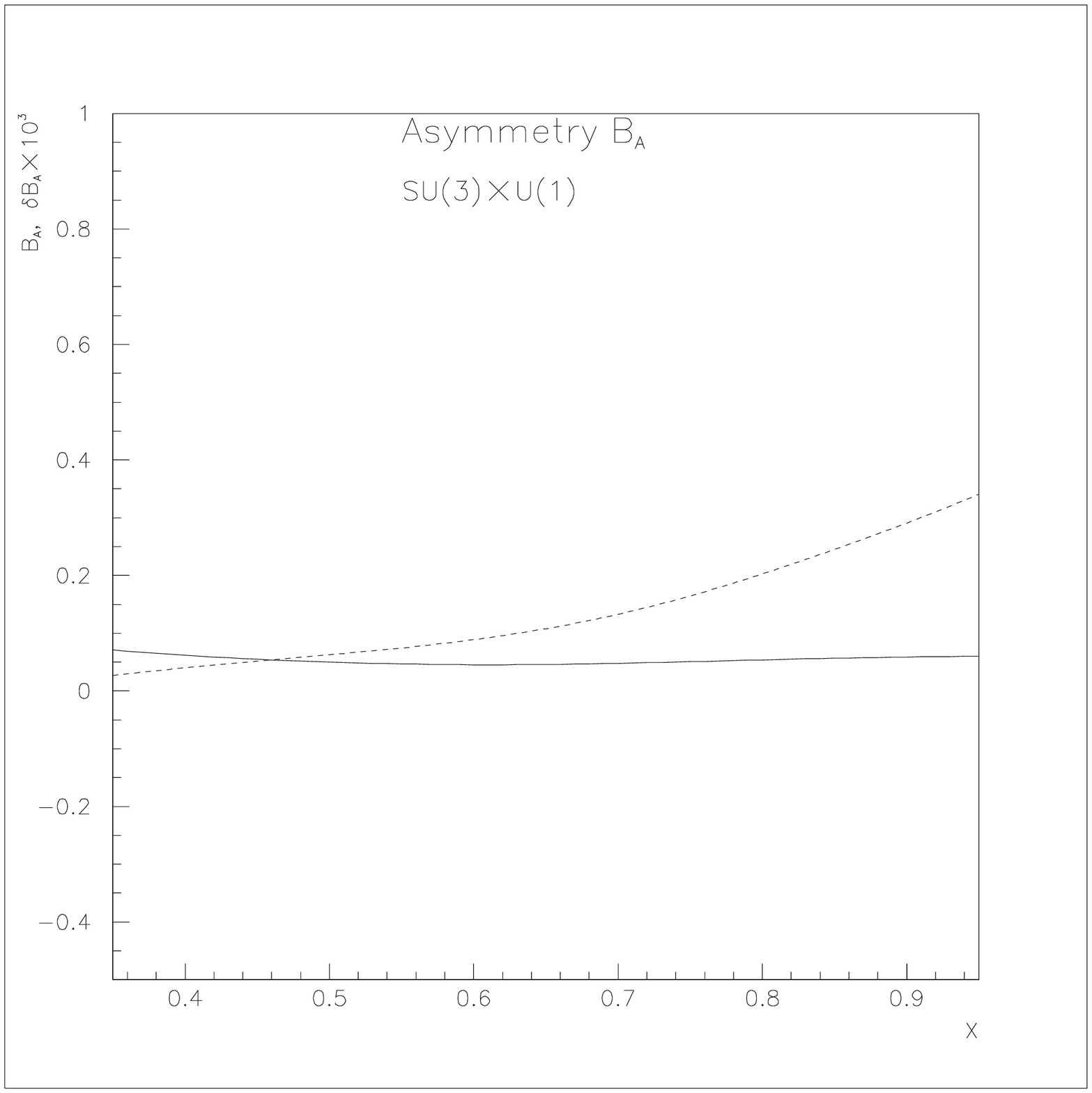}
\caption{The $B_{A}$ asymmetry in the Born approximation (solid line) and next to the Born approximation (dotted line)} \label{f9}
\end{minipage}\hfill
\begin{minipage}[b]{.475\linewidth}
\centering
\includegraphics[width=\linewidth, height=3.8in, angle=0]{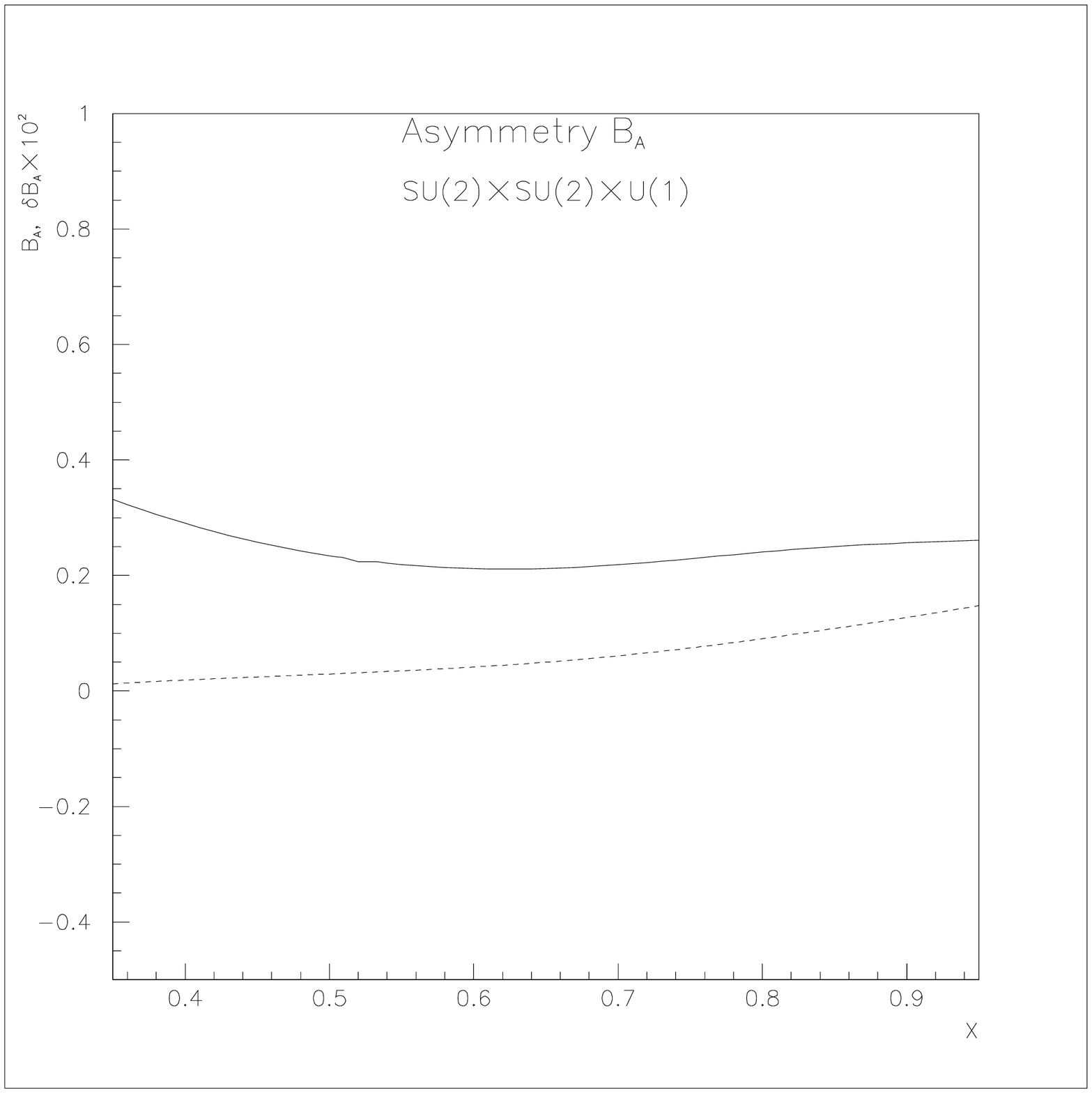}
\caption{The $B_{A}$ asymmetry in the Born approximation (solid line) and next to the Born approximation (dotted line)} \label{10}
\end{minipage}
\end{figure}
In spite of these facts inaccuracy defined by radiative
corrections it has insignificant values in the kinematic region for
nonminimal gauge models.

The investigation of the the electroweak asymmetries dependence
from the choice of the Weinberg angle revealed that the result
obtained in the Standard Model are slightly changed under the
$\sin^2\theta_W$ variation which is accorded by the experimental
data. The difference of asymmetries values under the same
$\sin^2\theta_W$ variation  in the nonminimal models can exceed $
10\%$ and at maximum x,y can become more than the asymmetries
values, with the sign of asymmetries changing  in the case
$A_{p,a}^{-}, C_{RR}, C_{RL}$.

\section{Conclusion}

$ $

The cross section for deep inelastic lepton-nucleon scattering
with arbitrary polarized initial fermions have been calculated within
the electroweak Standard Model and non-minimal gauge models with
additional neutral boson. The Lorentz-invariant formulas for cross
section and all types of electroweak asymmetries, polarized,
charge-polarized, charged, including the first-order electroweak
radiative correction have been obtained. We have presented a detailed numerical
discussion of the radiative corrections to different kinds of
electroweak asymmetries. The difference in the results minimal and nonminimal gauge models increases
significantly with the energy of incident leptons and reaches the
maximum for largest of scaling variable $y$,
where radiative corrections is important for precision
analysis of experimental data.

The  properties of the electroweak asymmetries considered here in
the different gauge models give the possibility  to verify the
Standard theory of the electroweak interaction and to define how
one can use the alternative gauge models. For these purposes the
analysis of the new experiment data from the DIS of the polarized
particles are used.

\section{Appendix}

$ $

All parameters for the discused nonminimal gauge model are peresented in the following
tables.
\\
\\
Table 1. The coupling constant in $\it SU(2)\times U(1)$ and $\it SU(3)\times
U(1)$ models.\label{t1}
\\
\\
\begin{tabular}{|c|c|c|}
\hline $Parameters$ & $\it SU(2)\times U(1)\rm model$ & $\it SU(3)\times U(1)
\rm model$
\\
\hline$g_Z$ & $e/\sin{2\theta_W}$ & $e/\sin{2\theta_W}$
\\
\hline$g_1^{'}$ & $1$ & $\cos{\alpha}$
\\
\hline$g_V^1$ & $-1/2 + 2\sin{\theta_W}^2$ & $-1/2 +
2\sin{\theta_W}^2 - A(-1/2 + 3/2\tan{\theta_W}^2)$
\\
\hline$g_A^1$ & $-1/2$ & $-1/2 + A(-3/2 + 1/2\tan{\theta_W}^2)$
\\
\hline$Q_{u,V}^1$ & $1/2(1-8/3\sin{\theta_W}^2)$ &
$1/2(1-8/3\sin{\theta_W}^2) - A(1/2 -5/6\tan{\theta_W}^2)$
\\
\hline$Q_{u,A}^1$ & $-1/2$ & $-1/2 - A(-1/2 -
1/2\tan{\theta_W}^2)$
\\
\hline$Q_{d,V}^1$ & $1/2(-1+4/3\sin{\theta_W}^2)$ &
$1/2(-1+4/3\sin{\theta_W}^2) - A(-1/2 +1/6\tan{\theta_W}^2)$
\\
\hline$Q_{d,A}^1$ & $1/2$ & $1/2 - A(-3/2 +1/2\tan{\theta_W}^2)$
\\
\hline
\end{tabular}
\newpage
\noindent
\begin{tabular}{|c|c|c|}
\hline $Parameters$ & $\it SU(2)\times U(1)\rm model$ & $\it SU(3)\times U(1)
\rm model$
\\
\hline$g_2^{'}$ & $...$ & $\sin{\alpha}$
\\
\hline$A$ & $...$ & $
\tan{\alpha}\cos^2{\theta_W}/(3-4\sin{\theta_W}^2)^{1/2}$
\\
\hline$g_V^2$ & $...$ & $-1/2 + 2\sin{\theta_W}^2 +B(-1/2 +
3/2\tan{\theta_W}^2)$
\\
\hline$g_A^2$ & $...$ & $-1/2 -B(-3/2 + 1/2\tan{\theta_W}^2)$
\\
\hline$Q_{u,V}^2$ & $...$ & $1/2(1-8/3\sin{\theta_W}^2) +B(1/2
-5/6\tan{\theta_W}^2)$
\\
\hline$Q_{u,A}^2$ & $...$ & $-1/2 +B(-1/2 - 1/2\tan{\theta_W}^2)$
\\
\hline$Q_{d,V}^2$ & $...$ & $1/2(-1+4/3\sin{\theta_W}^2) +B(-1/2
+1/6\tan{\theta_W}^2)$
\\
\hline$Q_{d,A}^2$ & $...$ & $1/2 +B(-3/2 +1/2\tan{\theta_W}^2)$
\\
\hline$B$ & $...$ & $
\cot{\alpha}\cos^2{\theta_W}/(3-4\sin{\theta_W}^2)^{1/2}$
\\
\hline
\end{tabular}
\\
\\
\\
Table 2. The coupling constants in $\it SU(2)\times SU(2)\times
U(1)\times U(1)$ and $\it SU(2)\times SU(2)\times U(1)$ models. \label{t2}
\\
\\
\begin{tabular}{|c|c|c|}
\hline $Parameters$ & $\it SU(2)\times SU(2)\times$ & $\it SU(2)\times SU(2)\times U(1) \rm model$
\\
 &  $\it U(1)\times U(1)\rm model$  &
\\
\hline$g_Z$ & $...$ & $e/\sin{2\theta_W}$
\\
\hline$g_1^{'}$ & $e/2$ & $\cos{\alpha}$
\\
\hline$g_V^1$ & $0$ & $-1/2 + 2\sin{\theta_W}^2 - A(-1/2 +
3/2\tan{\theta_W}^2)$
\\
\hline$g_A^1$ & $1$ & $-1/2 + A(-1/2 + 1/2\tan{\theta_W}^2)$
\\
\hline$Q_{u,V}^1$ & $0$ & $1/2(1-8/3\sin{\theta_W}^2) - A(1/2
-5/6\tan{\theta_W}^2)$
\\
\hline$Q_{u,A}^1$ & $-2/3$ & $-1/2 - A(1/2 - 1/2\tan{\theta_W}^2)$
\\
\hline$Q_{d,V}^1$ & $0$ & $1/2(-1+4/3\sin{\theta_W}^2) - A(-1/2
+1/6\tan{\theta_W}^2)$
\\
\hline$Q_{d,A}^1$ & $1/3$ & $1/2 - A(-1/2 +1/2\tan{\theta_W}^2)$
\\
\hline$g_2^{'}$ &
$\displaystyle\frac{e}{2\sin{\theta_W}\sqrt{1-2\sin{\theta_W}^2}}$ & $\sin{\alpha}$
\\
\hline$A$ & $
\displaystyle\frac{\tan{\alpha}\cos^2{\theta_W}}{(3-4\sin{\theta_W}^2)^{1/2}}$
 & $
\displaystyle\frac{\tan{\alpha}\cos^2{\theta_W}}{(3-4\sin{\theta_W}^2)^{1/2}}$
\\
\hline$g_V^2$ & $-1/2 + 2\sin{\theta_W}^2$ & $-1/2 +
2\sin{\theta_W}^2 +B(-1/2 + 3/2\tan{\theta_W}^2)$
\\
\hline$g_A^2$ & $-1/2 + 2\sin{\theta_W}^2$ & $-1/2 -B(-1/2 +
1/2\tan{\theta_W}^2)$
\\
\hline$Q_{u,V}^2$ & $1/2 - 4/3\sin{\theta_W}^2$ &
$1/2(1-8/3\sin{\theta_W}^2) +B(1/2 -5/6\tan{\theta_W}^2)$
\\
\hline$Q_{u,A}^2$ & $1/2 - 4/3\sin{\theta_W}^2$ & $-1/2 +B(1/2 -
1/2\tan{\theta_W}^2)$
\\
\hline$Q_{d,V}^2$ & $-1/2+2/3\sin{\theta_W}^2$ &
$1/2(-1+4/3\sin{\theta_W}^2) +B(-1/2 +1/6\tan{\theta_W}^2)$
\\
\hline$Q_{d,A}^2$ & $-1/2+2/3\sin{\theta_W}^2$ & $1/2 +B(-1/2
+1/2\tan{\theta_W}^2)$
\\
\hline$B$ & $
\displaystyle\frac{\cot{\alpha}\cos^2{\theta_W}}{(3-4\sin{\theta_W}^2)^{1/2}}$
& $
\displaystyle\frac{\cot{\alpha}\cos^2{\theta_W}}{(3-4\sin{\theta_W}^2)^{1/2}}$
\\
\hline
\end{tabular}

\end{document}